# User centred method to design a platform to design augmentative and alternative communication assistive technologies

Frédéric Vella[1], Flavien Clastres-Babou[1], Nadine Vigouroux[1], Philippe Truillet[1], Charline Calmels[2], Caroline Mercadier[2], Karine Gigaud[2], Margot Issanchou[2], Kristina Gourinovitch[2], Anne Garaix[2]

[1] IRIT Laboratory, Paul Sabatier University, 118 Route de Narbonne, 31 062 Toulouse Cedex 09, France
[2] OPTEO Foundation, 82 Rte de Saint-Mayme, 12850 Onet-le-Château, France
`Frederic.Vella@irit.fr`

**Abstract.** We describe a co-design approach to design the online WebSoKeyTo used to design AAC. This co-design was carried out between a team of therapists and a team of human-computer interaction researchers. Our approach begins with the use and evaluation of an existing SoKeyTo AAC design application. This step was essential in the awareness and definition of the needs by the therapists and in the understanding of the poor usability scores of SoKeyTo by the researchers. We then describe the various phases (focus group, brainstorming, prototyping) with the co-design choices retained. An evaluation of WebSoKeyTo is in progress.

**Keywords.** User centered design, co-design, AAC, multi-disabled person,

## 1 Introduction

Assistive technologies for communication and home automation allow people with disabilities to be autonomous and better social participation. However, many of these assistive technologies are abandoned [1] because they do not sufficiently take into account the expression of the needs of these people. In order for these technologies to meet the needs, it is important to involve, in user-centred design approach, occupational therapists and psychologists who can complement or express the needs of people with disabilities as part of their ecosystem [2]. In the field of augmentative alternative communication (AAC), their expertise allows to evaluate the abilities of the disabled person to better select and adapt the AAC. Some AAC sometimes integrate customization functionalities such as the communication board generator [3] or the Yellow Customize application [4], which allows the creation of one's own communication notebook. Currently, occupational therapists and psychologists use these functionalities for adapting and personalizing AAC.

In the framework of a research collaboration between therapists and researchers in human-computer interaction where the AAC were designed by IRIT researchers with the SoKeyTo platform [5], the therapists of the OPTEO Foundation expressed their need to possess this tool in order to design and adapt the AACs themselves for people with multiple disabilities.

In this paper, we will first present a state of the art on AAC design platforms and we will situate the SoKeyTo platform in relation to this state of the art. We will then describe the therapists' training approach as well as the tools and results of the therapists' evaluation of the SoKeyTo platform. These evaluation results led us to design the WebSoKeyTo platform. We will then describe the co-design approach implemented (brainstorming, low-fidelity mock-ups, focus-group) as well as the design choices that guided the design of the WebSoKeyTo platform in terms of functionalities for the design of AAC.

## 2 SoKeyTo Plateform and therapists' needs

### 2.1 Description of the SoKeyTo plateform

Sauzin et al., [5] have developed SoKeyTo which is a platform for AAC and home automation control interfaces. With this platform it is possible to design any type of interface (pictogram-based communication interface [6], mathematical input editor [7] and environment control [8]). It includes psychophysical models (Fitts [9], Hick-Hyman [10], [11] and Card [12]) for people with motor impairments that provide indicators of whether the interface is suitable for these people profiles.

SoKeyTo was/is used by researchers in human-computer interaction for the design of augmentative and alternative communication interfaces [13]. A lot of back and forth between therapists and researchers were therefore necessary to design customized AAC. This platform is composed of two components: the editor component and the generation of an executable AAC with a configuration interface (type of interaction, feedback, scanning system parameters, etc.).

The editor component allows to define the morphology and the contents of the AAC buttons, the layout and the structure of the AAC, the visual and audio feedback and the type of associated functions (communication function, running an application, sending a message to be broadcast by a text-to-speech system). It also enables to associate to each button the communication protocol (MQTT : Message Queuing Telemetry Transport, [14], bus IVY [15], https, radio frequency, infrared) needed to interact with the connected objects or devices used by the disabled person.

The player component allows the customization of the AAC and the interface between several input interaction modes (pointing device, eye tracker, joystick, speech recognition, on/off switch). The platform also allows various control modes to be configured (pointing, time delay click, scanning system [16]) according to the abilities of disabled people.

**Table 1.** Application designed with the SoKeyTo platform.

| Type of applications | Example of applications |
|---|---|
| AAC [13] : Example of communication page : white background Communication button, yellow background, navigation color. | 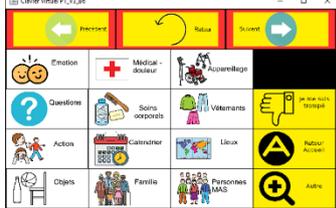 |
| Text input interface: Two illustrations: on the left, an input keyboard designated by the person with a disability herself [17] and on the right, a mathematical input keyboard [7]. | |
| Environment control interface [8]; This interface allows to control a home automation system (light, door, etc.). The SoKeyTo platform made it possible to take into account accessibility rules (maximisation of contrast: black background and white characters). | 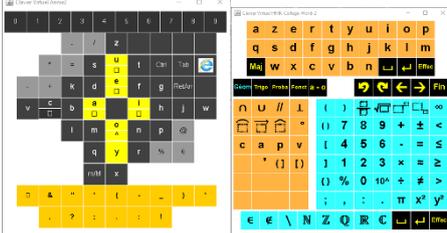 |
| Command interface of video games: Virtual keyboard of video game controllers. | 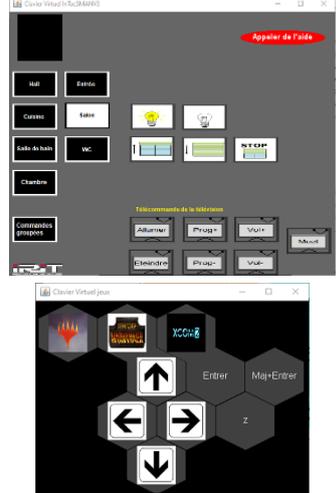 |

Table 1. gives 4 examples of applications designed by the SoKeyTo platform by researchers in human-computer interaction. The functionalities of SoKeyTo show that it is possible to design several types of input applications, navigation keyboards or even environment control applications accessible to the elderly. Designed and used by researchers, the question is whether this platform is efficient, useful and usable by expert AAC therapists? The ergonomic study carried out on the Bastien and Scapin criteria reveals some limitations: no und-o/redo function, little flexibility in the commands, no alternative commands, little feedback, no error handling.

## 3 Related work on AAC design platforms

Many AAC systems exist on the market for people with multiple impairments whose communication needs may evolve according to their human environment and abilities. The analysis of the functionalities of some AAC authoring applications (see Table 2) shows that most of them integrate editing and configuration functions to take into account these evolving needs of AAC. Table 2 lists the main features of these applications and comparatively the missing functionalities of SoKeyTo.

Table 2. Applications in competition with SoKeyTo.

| Competing Applications | Features | Functions not available in SoKeyTo |
|---|---|---|
| EDiTH [16] | Customisable AAC interface; button customization; interface model availability; column scan strategy; | Access to the different interfaces through menus ; |
| Pictocom [18] | AAC Interface Editor; several interaction modes; integrated home automation extension; execution of software on a computer | Combination of several modes on one button (text, sound, image); |
| Boardmaker7 [19] | AAC Interface Editor; execution of software on a computer; predefined page templates ; several interactions modes ; | Teaching oriented software; Printing of AAC ; |
| AraBoard [20] | AAC Interface Editor and Player ; customization ; on line web application ; promote Arasaac symbols [21]; | Usable under Android; export of the AAC in pdf ; |
| Cboard [3] | AAC Interface Editor ; on line web application ; | Usable under Android ; History of pictographic communication in a digital tape; Printing of AAC ; Multilingual ; |
| Jellow [4] | AAC Interface Editor ; on line web application ; customization ; | Extension to uses for the hearing impaired; Multilingual ; |

| Proloquo2Go [22] | AAC Interface Editor ; predefined page templates ; import of pictograms ; customization, execution of software on a computer | Usable under Mac Os ; Creation of a customized synthetic speech; History of pictographic communication in a digital tape; |

All these platforms integrate an editor and a player. The editing features differ between the applications. Two of them offer the possibility to use page templates [19], [22] and two others offer the possibility to have a band for the communication history [22], [3]. Several platforms [5], [18], [19] allow interfacing with several control modes. The PictoCom platform [18] is the closest to the SoKeyTo platform because it includes the home automation and communication aspect in the AAC. A strong point is the ability of SoKeyTo to configure several scanning strategies [13] and to allow to associate several communication protocols to a button. This allows a priori to control many connected objects or applications that a person with a disability could use to increase his autonomy. This comparative analysis allows to identify and discuss the interest of having these missing functionalities in the SoKeyTo.

The objective of the article is to describe the process implemented to evaluate the usefulness and usability of SoKeyTo and the co-design of the new Web-SoKeyTo platform.

## 4  Co-Design of the WebSoKeyTo platform using a user-centred design method

### 4.1  Therapists' needs

The transition from SoKeyTo to WebSoKeyTo is the result of a request from therapists and characteristics analysis of the applications listed in table 2. Indeed, the researchers designed the AAC based on the needs and specifications of the AAC provided by the therapists. However, this collaboration has shown its limits:

— The design of the AAC was time-consuming, with various back and forth between the two parties (researchers/therapists); these delays had negative consequences: the AAC was no longer suitable for the abilities of the multi-disabled person and/or the new needs were taken into account too late;
— The non-adaptation of AAC during the therapists' sessions: if the therapist wished to make adaptations during the appropriation or use sessions these were not possible.

These two reasons made the therapists wish to have autonomy in the design of AAC and thus to have an increased efficiency in its implementation. The ergonomic tests of Bastien and Scapin [17] and the needs of therapists for AAC tools led us to:

— Firstly, to carry out a user experience of SoKeyTo;

– Secondly, to implement a user-centred design method for the design of WebSoKeyTo accessible to therapists for the design of AAC adapted to the needs of people with disabilities.

### 4.2 Co-design tools of the WebSoKeyTo platform

Co-design uses a collaborative team approach that allows non-designers to become equal members of the design team. Sanders and Stappers [23] defined "*Co-design is a specific instance of co-creation practice that allows users to become part of the design team as 'experts of their experience*". "*It represents a shift away from design as the task of individual experts towards using the collective creativity of a team with members from different backgrounds and interests*" [24]. We deployed this co-design approach in the design of the WebSoKeyTo platform which has involved two type of therapists (psychologist and occupational therapists) and researchers in human-computer interaction. Table 2 shows the main stages of this process. In order to measure the contribution of the co-design of WebSoKeyTo, we first conducted an evaluation of the SoKeyTo platform which will constitute a baseline evaluation.

**Table 3.** Steps of the co-design of WebSoKeyTo.

| Platform | Steps | Outcomes |
|---|---|---|
| SoKeyTo | Training and trials during two months | Experience in designing AAC |
| | User Experience | UX Value (USE [26] and AttrakDiff [27] and verbatim of open questions |
| | Focus Group | Therapists' needs |
| WebSoKeyTo | Brainstorming | Functional and ergonomic requirements |
| | Prototyping | Middle mock-ups and high fidelity prototype |
| | User Experience | UX Value (USE and AttrakDiff ) |

### 4.3 Therapist population

Six therapists (2 occupational therapists and 4 psychologists) were recruited to participate in the user centred approach to co-design WebSoKeyTo. Five therapists had never used another AAC design tool before and one occupational therapist is expert in the use of SoKeyTo.

### 4.4 Assessment of SoKeyTo

**Training of SoKeyTo**
Firsly, we trained the therapists to use the SoKeyTo platform by demonstrating all the features in a practical way They were invited to use the SoKeyTo platform for two months: firstly, a scenario imposed by the SoKeyTo platform designers for one month, and then a free scenario for the design of an AAC for a disabled person. These therapists could benefit from the help of the SoKeyTo designers in case of bugs or difficulties of use. At the end of this trial phase of the SoKeyTo platform, we proceeded to the evaluation of the usability of this platform by means of the USE (Usefulness, Satisfaction, and Ease of use) questionnaire [26], the AttrakDiff [27] and open questions (failures, most useful functionality, missing functionality and estimated usefulness). They reported the difficulties in ease of use, low satisfaction with the function mode and average learning and usefulness [28]. The therapists mentioned that the SoKeyTo platform were too computer-oriented and not needs-oriented enough.

**User Experience**
Figure 1 shows the results of the AttrakDiff questionnaire [27]. We have chosen not to average the results by type of therapist, given the very small sample size. However, we believe that the individual results are of interest to the platform's designers. Indeed, the AttrakDiff questionnaire quantifies both the pragmatic qualities (perceived usefulness and usability) and the hedonic qualities (subjective emotions felt) of a digital system. This measurement of hedonic qualities provides a significant added value when we want to understand how our users feel. The results show that for 4 therapists (two occupational therapists and two psychologists) the results are neutral for both axes (-1, +1). One psychologist [-0,2, 4] characterizes the hedonic quality of the platform more positively. These results show that satisfaction is minimal and that improvements need to be made. The evaluation of the computer psychologist is in the superfluous zone (-3,- 1) for both axes. The use of the SoKeyTo platform has therefore generated dissatisfaction for this user. This evaluation shows that the usability and hedonic qualities must be improved.

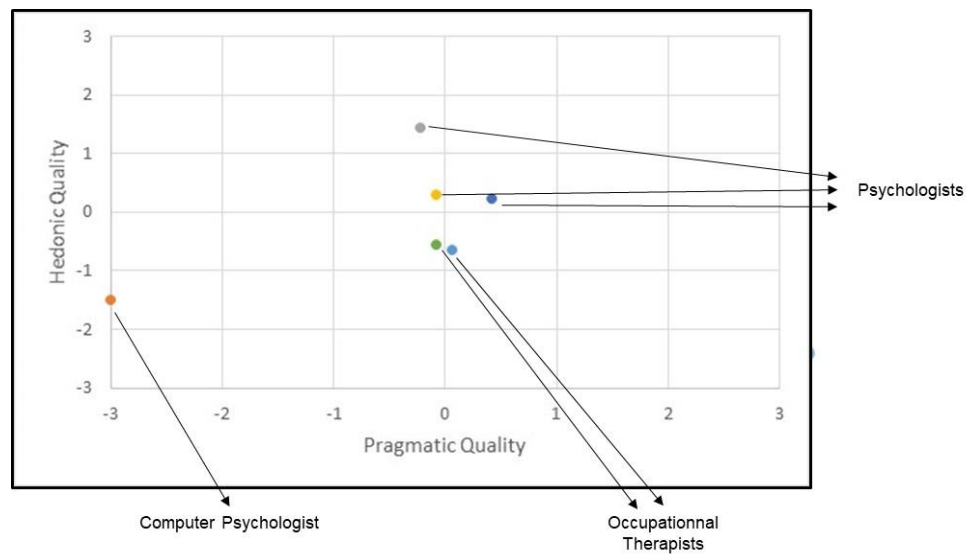

**Fig. 1.** AttrakDiff results (Portfolio representation).

We complemented this evaluation with 4 open questions in order to identify limitations that would explain the poor scores of the AttrakDiff scale and that could increase usefulness and usability. Table 4 lists the topics of these questions (identified limitations, most useful and missing features, perceived usefulness).

Table 4. Open questions to therapists.

| Questions | Summary of therapists' answers |
| --- | --- |
| Q1: Have you encountered any problems in using the SoKeyTo platform? Which ones? | Bugs; Lack of visual feedback; Lack of un-do/redo; Logic problems (management of pictograms and pages; bad logic for accessing the functions of creating and modifying an AAC button); No consideration of the screen resolution for the display of the AAC; Complexity in the parameterization of the scanning technique; |
| Q2: What are the existing functionalities used to design interfaces? List them | Link between pages; ability to associate actions with keys; Platform-independent execution of AAC; Choice of buttons characteristics; |
| Q3: What are the missing features that are essential to design in the new platform? List them | Access to databases of pictograms or pages of pictograms (read and write); Overview of links between pages; Independence between the editor and runtime components of SoKeyTo; Generation of the size of the keys according to the screen resolution; undo/redo function; Compatibility with the principles of Windows for the functions accessible by right-click; More ergonomic setting of the scanning; History of the pictogram communication; |
| Q4: Is this platform an indispensable tool for the complete execution of your professional activity? | Yes need to have this kind of tool, as it is more adaptable than existing applications; Yes, to meet the customization needs of AACs; Not essential as it stands, but the idea of developing this tool is very interesting and novel. |

The end of the testing phase highlighted therapists' frustrations with the functional limitations and bugs of the SoKeyTo platform. These limitations were a hindrance to learning and using the platform. Indeed, for question Q2, three of the therapists did not answer because of the lack of use. Four of the six therapists report that the platform will help meet the needs of their patients in designing AAC. The other two therapists have negative opinions of the current state of the platform due to the bugs and functional limitations reported in questions Q1 and Q3 respectively.

The analyses of the SoKeyTo evaluation show the usefulness of such a tool but the need to design a more ergonomic AAC design platform that better meets the needs of therapists.

**Focus Group**

A focus group between the six therapists and three researchers in human-computer interaction enabled us to define the priority needs of an AAC design platform. Therapists have taken an active role in making conceptual artifacts via function proposal card and and audio explanations based on their digital experience that express ideas hox they wish to use the WebSoKeyTo platform. These needs are mainly listed in Table 4 (answers Q1 and Q3).

### 4.5 WebSoKeyTo Design Process

**Brainstorming**

The people who participated in this brainstorming are 3 seniors and two students in human-machine interaction. One of the participants is the developer of the SoKeyTo application. The objective of this brainstorming was to think about proposing a more ergonomic, attractive and fun interface. The proposals focused on handling interaction techniques (management of zoom in page navigation, proposal for shortcuts and affording icons), visual and sound feedback to be associated with the buttons and the assistance mechanisms to be implemented. These proposals were used for the medium fidelity mock-up.

**Medium fidelity mock-ups**

The mock-ups were developed based on the results of the design team's brainstorming and the needs expressed and discussed in the focus group by the therapists. We present three interface mock-ups (specification of an AAC button, page navigation, and specification of scanning strategies). The mock-ups were developed with AdobeXD or Balsamiq.

*Specification interface*

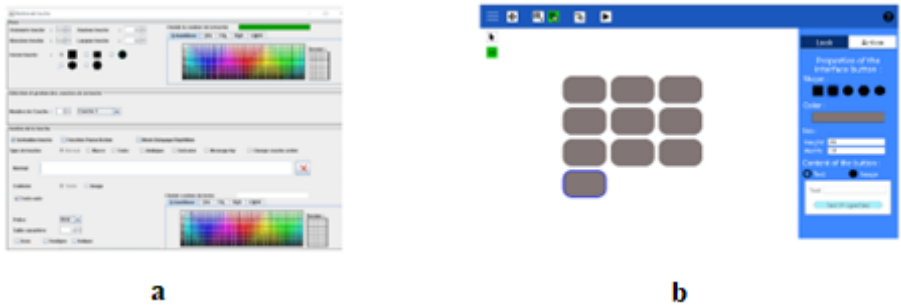

**Fig. 2.** Specification interface: a. SoKeyto ; b. WebSoKeyTo.

In the SoKeyTo, the interface for specifying characteristics (morphological and associated actions) was superimposed on the key design grid (Figure 2.a). An ergonomic study using the Bastien and Scapin criteria [25] confirmed that the presence of a side panel (Figure 2.b) was more accessible and allowed feedback on the effects of key creation and modification actions. Work on structuring and naming items was also proposed (morphological characteristics, specification of actions associated with keys). Examples of actions were proposed and completed by the therapists (dependency link, access to applications such as music players, etc.). The therapists validated the mock-up.

*Page navigation.*

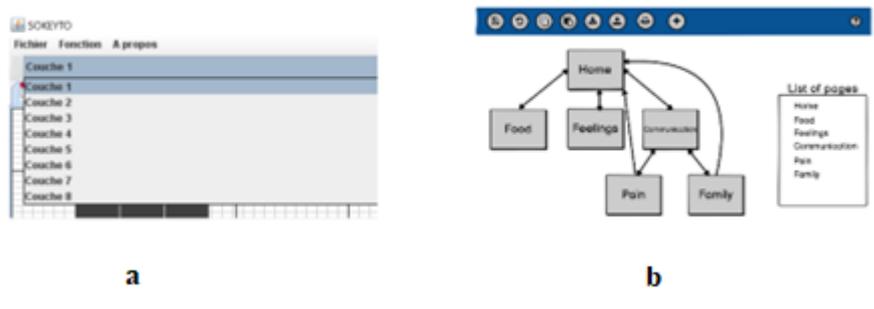

**Fig. 3.** Page Navigation: a. SoKeyTo; b. WebSoKeyTo.

The way of handling pages with SoKeyTo posed a problem: access to pages in list form, no representation of links between pages (Figure 3.a). The representation in the form of a graph (Figure 3.b) was proposed to the therapists in order to visualize the dependency links between pages. This visual representation is associated with the list of pages. As the number of pages for a AAC can reach more than 50 pages [13], the question of representation and handling will have to be addressed by evaluating the techniques of Fisheye tree views and lenses for graph visualization [29] and of tree map [30] which consists of dividing the interface into several zones by giving more space to the focus.

*Scanning strategy.*

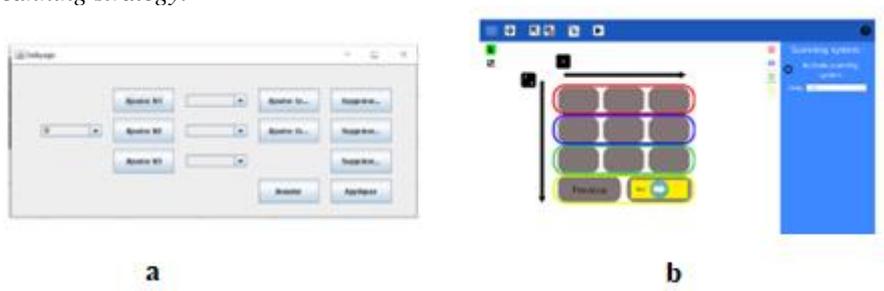

**Fig. 4.** Scanning strategy: a. SoKeyTo; b. WebSoKeyTo

The SoKeyTo platform makes it possible to define several interaction modes, including the scanning technique [16]. The configuration parameters are the scanning direction

(row then column or column then row, the specification of the scanning block, the scanning speed, the associated visual feedbacks, etc.), the scanning speed, the scanning direction and the visual feedbacks. The SoKeyTo specification interface (Figure 4.a) was found to be very difficult for therapists to understand for specifying scanning strategies via a list of buttons. The mock-up (Figure 4.b) proposes visual elements to indicate scanning direction (horizontal and vertical arrow) and order (dice symbol 1 and 2). These more affording proposals have been validated to set the scanning order.

## 5 Discussion of the co-design approach

We implemented a co-design approach between a team of human-computer interaction researchers and a team of therapists. We started with a training session on the SoKeyTo platform, which allowed the therapists to discover the functionalities for designing an AAC. This training phase and the two months of use allowed the therapists to identify their needs and to propose new functionalities, essential in their professional activity. However, the evaluation of SoKeyTo's negative pragmatic and hedonic quality scores and the needs expressed highlighted the need: 1) to design a complete, ergonomic tool for designing AAC online and oriented to the therapists' profession; 2) to implement a co-design. Except for the SoKeyTo training, which took place face-to-face, all the co-design activities (1 focus group on needs, two brainstormings on low-fidelity mockups, 3 focus groups following the medium-fidelity models) took place by videoconference due to the sanitary conditions of the COVID. The collaborative work and video-conferencing tools allowed for numerous and fruitful exchanges during the focus group to identify the needs and those to present the mock-ups. Arbitration took place on divergent points of view by majority consensus. This approach will be continued for the functionalities (pictogram editor and access to the page and pictogram databases). Another extension is the creation of importable page templates and page categories between therapists to increase the speed of design of AAC. The WebSoKeyTo application is online and used by therapists to design AACs and a new user experience is planned to qualify the benefits of the redesign of SoKeyTo.

## 6 Conclusion

Firstly, we performed a state of the art on AAC design platforms and we situated the SoKeyTo platform in relation to this state of the art. Then, we describe a co-design approach to design the online WebSoKeyTo used to design AAC. This co-design was carried out between a team of therapists and a team of human-computer interaction researchers (HCI). Our approach begins with the use and evaluation of an existing SoKeyTo AAC design application. This step was essential in the awareness and definition of the needs by the therapists and in the understanding of the poor usability scores of SoKeyTo by the researchers. Numerous exchanges where both knowledge (AAC profession and HCI) took place during the numerous focus groups and brainstorming to identify the needs and validate the mockups. This co-design approach highlighted that the expression of needs and the solution evolved together as reported by Sanders and Stappers [23]. As a perspective, we plan to: 1) analyze the frequency of use of WebSoKeyTo functions by therapists; 2) measure the design and adjustment time of a AAC, 3) analyze the impact of WebSoKeyTo use on therapists' activities and relationships with caregivers; 4) propose the WebSoKeyTo interface according to the context of action in the next versions through adaptation algorithms that would be deduced from activity logs.

## 7      Acknowledgment

This work is partially funded by the ANS (Agence du Numérique en Santé, France) in the framework of Structures 3.0 program.